\documentstyle[12pt]{article}
\begin{document}
{\hskip 12.0cm} AS-ITP-95-06\\
\vspace{1ex}
{\hskip 12.5cm} February, 1995\\
\vspace{5ex}
\begin{center}
{\LARGE Quark Mass Hierarchy and CP Violation in Low Energy Supersymmetry}\\
\vspace{4ex}
{\sc Chun Liu}\\
\vspace{2ex}
{\it  CCAST (World Laboratory) P.O. Box 8730, Beijing, 100080}\\
and
{\it  Institute of Theoretical Physics, Academia Sinica}\\
  {\it P.O. Box  2735,  Beijing  100080, China
\footnote{Mailing address}}\\

\vspace{5.0ex}
{\large \bf Abstract}\\
\vspace{3ex}
\begin{minipage}{130mm}

   A pattern of quark mass hierarchy and CP violation within the
framework of low energy supersymmetry is described.  By assuming some
discrete symmetry among the three families, the quarks of the third
family obtain masses at tree level.  The second family obtains
masses radiatively at one-loop level due to the
soft breaking of the family symmetry.  At this level, the first family
remains massless by some degeneracy conditions of the squarks.  As a
result of R-parity violation, the sneutrino vacuum expectation values
are nonvanishing.  CP violation occurs through the superweak sneutrino
exchange.  This picture is consistent with the experiments on the
flavor changing neutral current.\\
\end{minipage}
\end{center}

\newpage

{\large \bf 1. Introduction}\par
\vspace{1.0cm}
   The origin of the fermion mass hierarchy as well as the CP violation
remains one of the most important problems in the Standard Model.  The
solution for this problem must lie in some physics beyond the Standard
Model.  Up to now, the most favorable framework which is beyond the
Standard Model is supersymmetry, because it can help to solve the gauge
hierarchy problem.  Therefore it is natural to study the origin of the
fermion mass hierarchy and the CP violation within the framework of
supersymmetry.  However, there are still two different kinds of ideas
on this subject.  One is to relate this origin with the ultrahigh
energy physics, and another with the low energy physics at the scale
of about 1TeV.\par
\vspace{1.0cm}
   In addition to supersymmetry, it is important to assume some
horizontal symmetry among the three families to understand the pattern
of the fermion mass hierarchy.  Usually such assumption in the low
energy physics is more necessary than that in the high energy physics.
This horizontal symmetry has to be violated in order to give the
realistic fermion spectra.  Different from introducing other new
particles, the violation can be made by using the properties of the
superpartners of the known particles.  Strictly speaking, if the
three sneutrinos have different vacuum expectation values (VEVs)
after electroweak breaking, which are the particles of Higgs-like
except that they carry lepton number, the family symmetry can be
broken.  Furthermore, this family symmetry breaking can be also
introduced explicitly in the soft sector, as in the case of
supersymmetry breaking.\par
\vspace{1.0cm}
   Within the framework of low energy supersymmetry with  R-parity
violation [1] and family symmetry [2], the lepton mass hierarchy has
been studied in Ref. [3].  By assuming a cyclic symmetry among the
left-handed doublets of the three families of leptons, the
hierarchical pattern can be obtained naturally.  The tau lepton gets
its mass from the VEV of the Higgs field, the muon gets its mass from
the VEVs of the sneutrino fields, whereas the electron is massless
at tree level and obtains mass from loop level due to the soft breaking
of the family symmetry.\par
\vspace{1.0cm}
   In this paper, we study the quark mass hierarchy and the CP violation
in the framework of low energy supersymmetry, especially in which
with R-parity violation.  We also assume some discrete family symmetry
which breaks softly.  The third family of quarks gets masses from the
Higgs fields.  There are two Higgs doublets, one couples to the up
quarks, and another to the down quarks.  The large ratio of the top
quark mass to the bottom quark mass is partly explained by the different
VEVs of the Higgs fields.  Because of the family symmetry assumed, the
other two families do not obtain mass from the tree level.  This is
different from the lepton case of Ref. [3].  These two families can
obtain their masses from the loop level.  The supersymmetric radiative
mass generation of quarks and leptons were pointed out in Refs. [4] and
[5], respectively.  Quark mass  gets a large color contribution which is
absent for lepton mass.  So the quark is generally heavier than the
corresponding lepton.  Hence the radiative mass generation is viable.
We further find out the way to obtain the
hierarchy of the second and the first families.  Consequently, the
quark mixing matrix is discussed.\par
\vspace{1.0cm}
   To bring our discussion to completion, the CP violation has to be
considered.  It is at this stage that the R-parity violation plays its
essential role.  To keep the proton off rapid decay, the baryon number
conservation is adopted.  The sneutrinos have the same quantum numbers
as one Higgs field except for their lepton number which is not conserved
however.  Generally they can get nonvanishing complex VEVs.  This is
similar to the multi-Higgs doublets models.  Because of the
phenomenological requirements from the lepton universality, the sneutrino
VEVs should be much smaller than the Higgs VEVs.  In this case, we would
like to point out that the CP violation can occur through the sneutrino
exchange.\par
\vspace{1.0cm}
   This paper is organized along the above outline as follows.  In
Sec. 2, our framework of low energy supersymmetry with R-parity
violation and family symmetry is described.  In Sec. 3, the quark mass
hierarchy and mixing are studied.  In Sec. 4, the CP violation is
considered.  The constraints to the squark masses from the flavor changing
neutral currents (FCNC) are analyzed in Sec. 5.  We summarize and discuss
our results in the final section.\par
\vspace{1.5cm}
{\large \bf 2. The Framework}\par
\vspace{1.0cm}
   The low energy supersymmetric models with R-parity violation [1] is
one kind of natural, supersymmetric extensions of the Standard Model.
R-parity is a multiplicative parity which is defined to be +1 for the
ordinary particles and $-1$ for superpartners.  In addition to the
supersymmetric gauge interactions and supersymmetric Yukawa interactions
which conserve R-parity, the gauge invariance allows the following
R-parity violating interactions in the low energy supersymmetric models,
\begin{equation}
f_{\Delta L}=\lambda_{ijk}L_iE_j^cL_k+\lambda'_{ijk}Q_iD_j^cL_k~,
\end{equation}
and
\begin{equation}
f_{\Delta B}=\lambda_{ijk}''U_i^cD_j^cD_k^c~,
\end{equation}
where $i$, $j$, $k$ denote the three families, the left-handed chiral
superfields $L$ and $Q$ correspond to the SU(2) doublet lepton and quark
fields respectively, $E^c$, $U^c$ and $D^c$ correspond to the SU(2)
singlet antiparticle fields of leptons, up-type quarks and down-type quarks
respectively.  Eq. (1)
breaks lepton number and Eq. (2) breaks baryon number invariance.  To keep
the proton off rapid decay, some restrictions beyond gauge invariance
should be adopted, e.g., the well-known R-parity invariance which,
however, has no more natural motivation than some other alternative
choices if they are also at experimentally acceptable level.  We will adopt
the baryon number invariance.  For some specific structure of family indices,
the R-parity violating interactions with baryon number conservation can be
consistent with both the laboratory [1] and the astrophysical experiments
[6].  It should be noted that in this case the R-parity  (or lepton number)
violating interaction (1) does not vanish.  However the form
of this interaction is still too arbitrary to give some definite predictions.
This arbitrariness can be reduced by introducing family symmetry.\par
\vspace{1.0cm}
   It is interesting if there is some symmetry among the three families.
The idea of family symmetry was used in the attempts to understand the
fermion mass hierarchy problem [2].  A discrete ${\rm Z_3}$ cyclic family
symmetry of the three left-handed lepton doublets was proposed in Ref. [3].
{}From the general fact that the third family of quarks is much heavier than
the other two families, the quark mass matrix of democratic mixing [7] is
usually assumed, with which only the third family of quarks is massive.
This mixing implies some family symmetry in the quark sector.  Of course,
such symmetry has to be slightly violated to make the other two families
of quarks become massive.  This happens if the family symmetry can be
broken softly.  So that the other two families of quarks get
masses radiatively [4].  It should be noted that the choice of the family
symmetry corresponding to the democratic mixing is not unique.  For
simplicity, in addition to the ${\rm Z_3}$ cyclic family symmetry among
the left-handed doublets of leptons, which is denoted as ${\rm Z_{3L}}$,
we assume another ${\rm Z_3}$ cyclic family symmetry
which is among the three left-handed
doublets of quarks, and is denoted as ${\rm Z_{3Q}}$.\par
\vspace{1.0cm}
   Our framework in low energy supersymmetry is based on the above
adopted R-parity violation and assumed ${\rm Z_{3L}\times Z_{3Q}}$
family symmetry.  In this paper, we focus on the quark sector.  The
supersymmetric gauge interactions are uniquely determined and can be found
in text books.  With the left-chiral lepton superfields and their
${\rm SU(2)\times U(1)}$ quantum numbers $L_i(2,-1)$ and $E^c_i(1,2)$, the
quark superfields $Q_i(2,\frac{1}{3})$, $U^c_i(1,-\frac{4}{3})$ and
$D^c_i(1,\frac{2}{3})$, the Higgs superfields $H_u(2,1)$ and $H_d(2,-1)$,
the superpotential relevant to the quark sector of our model which
possesses the ${\rm SU(2)\times U(1)}$ gauge symmetry, the
${\rm Z_{3L}\times Z_{3Q}}$ discrete family symmetry and baryon number
conservation is
\begin{equation}
{\cal W}=g^u_k(\sum_{i}^{3}Q_i^a)H_u^bU_k^c\epsilon_{ab}
+g^d_k(\sum_{i}^{3}Q_i^a)H_d^bD_k^c\epsilon_{ab}
+\lambda'_k\sum_{i,j}^{3}(Q_i^aL_j^b)D^c_k\epsilon_{ab}~,
\end{equation}
with $a$ and $b$ being the SU(2) indices.  The first two terms are the
Yukawa interactions, their couplings only depend on the flavor of the SU(2)
singlet quarks.  In addition, the soft supersymmetric breaking terms which
include gaugino and scalar mass terms, as well as the trilinear scalar
interactions should be added in the Lagrangian.  We assume that the
${\rm Z_{3L}\times Z_{3Q}}$ discrete symmetry is violated in the soft
breaking terms so as to generate the light quark masses radiatively.
Therefore, the couplings of the soft breaking terms are still arbitrary in
general.  The constraints to them will be discussed from the aspects of
phenomenological analysis.\par
\vspace{1.0cm}
   The vacuum is determined by the scalar potential of this model in which
the Higgs sector has been also included [3].  The ${\rm SU(2)\times U(1)}$
gauge symmetry breaks down spontaneously.  The sneutrinos have the same
quantum numbers as the Higgs field $H_d$ except for the lepton number.
Because the lepton number is not conserved, the sneutrino fields play the
same role as the Higgs field.  Therefore both the Higgs fields and
the sneutrino fields obtain nonzero VEVs which can be determined by the
parameters of the scalar potential.  In this paper, instead of being
involved in the detailed analysis of the scalar potential, we leave the
values of the Higgs and sneutrino VEVs to be fixed from some
phenomenological constraints.  From the discussion of the lepton sector
in Ref. [3], the sneutrino VEVs are much smaller than the Higgs VEVs.  This
is required not only by the lepton universality, but also by the
explanation of the lepton mass hierarchy.  The implication of the
smallness of the sneutrino VEVs on the CP violation of the quark sector
will be analyzed later in this article.\par
\vspace{1.5cm}
{\large \bf 3. The quark mass hierarchy and mixing}\par
\vspace{1.0cm}
   The family symmetry gives a hierarchical pattern to quark masses.
After the ${\rm SU(2)\times U(1)}$ gauge symmetry breaking, the
superpotential (3) makes the quarks become massive.  By denoting the VEVs
of Higgs fields $H_u$, $H_d$ and sneutrino fields $L_i$ ($i$=1, 2, 3) as
$v_u$, $v_d$ and $v_i$, the tree level mass matrix of the up quarks
is
\begin{equation}
M^u=\left( \begin{array}{ccc}
g_1^uv_u&g_2^uv_u&g_3^uv_u\\
g_1^uv_u&g_2^uv_u&g_3^uv_u\\
g_1^uv_u&g_2^uv_u&g_3^uv_u\\ \end{array} \right)~;
\end{equation}
the tree level mass matrix of the down quarks is
\begin{equation}
M^d=\left( \begin{array}{ccc}
g_1^dv_d+\lambda'_1\sum_{i}^{}v_i&g_2^dv_d+\lambda'_2\sum_{i}^{}v_i&
g_3^dv_d+\lambda'_3\sum_{i}^{}v_i\\
g_1^dv_d+\lambda'_1\sum_{i}^{}v_i&g_2^dv_d+\lambda'_2\sum_{i}^{}v_i&
g_3^dv_d+\lambda'_3\sum_{i}^{}v_i\\
g_1^dv_d+\lambda'_1\sum_{i}^{}v_i&g_2^dv_d+\lambda'_2\sum_{i}^{}v_i&
g_3^dv_d+\lambda'_3\sum_{i}^{}v_i\\ \end{array} \right)~.
\end{equation}\par
\vspace{1.0cm}
   At tree level, the up quark mass matrix originates from the Yukawa
interactions only.  However the down quark mass matrix involves the
contributions of both the Yukawa and the R-parity violating interactions.
Although the difference in the values of the sneutrino VEVs breaks the
${\rm Z_{3L}}$ symmetry, it does not affect the ${\rm Z_{3Q}}$ symmetry.
Both the up quark and the down quark mass matrices exhibit the
${\rm Z_{3Q}}$ symmetry explicitly.  They are of rank-one and can be
regarded as a kind of democratic family mixing [7].  Only one family,
the third family, is massive at tree level.  Because $v_i\ll v_d$, the
value of the bottom quark mass is not affected significantly by the
R-parity violating interactions.  The large ratio of the top quark mass
to the bottom quark mass $m_t/m_b$ is partly explained by the different
VEVs of the two Higgs fields.  Nevertheless, a hierarchy between the third
family and the other two families is resulted in.\par
\vspace{1.0cm}
   It is at the loop level that the masses of the other two families of
quarks are generated through the soft breaking of the ${\rm Z_{3Q}}$
symmetry.  The generation of the quark masses by supersymmetric radiative
corrections has been pointed out in Ref. [4].  Both the light up quark
and down quark masses can be induced naturally through the one-loop
diagram Fig. 1 in this model, where $\tilde{g}$ and $q$ stand for the
gluinos and the quarks, respectively.  The squark mass terms are assumed
to be flavor diagonal which will be justified through the analysis of FCNC
later.  The mixings of the scalar quarks associated with different
chiralities are due to the trilinear soft breaking terms.  The structures
of the mixings for the up squarks and the down squarks have the same form
as that shown in matrices (4) and (5).  These mixings are multiplied by
some common supersymmetric mass parameters $\tilde{m}$'s.  They can be
expressed as $\tilde{m}^u_jv_u$ and  $\tilde{m}^d_jv_d$ for the up squarks
and the down squarks, respectively, where $j$ is the flavor index of the
right-handed squark, and the contribution of the sneutrino VEVs has been
neglected.   Fig. 1 contributes to the quark mass matrices (4) and (5) the
following terms,
\begin{equation}
(\delta M)_{ij}=\frac{\alpha_s}{\pi}\frac{2m_{\tilde{g}}}
{m^2_{\tilde{g}}-m^2_{\tilde{q}^c_j}}(\frac{m^2_{\tilde{g}}}{m^2_{\tilde{g}}
-m^2_{\tilde{q}_i}} \ln \frac{m^2_{\tilde{q}_i}}{m^2_{\tilde{g}}}
+\frac{m^2_{\tilde{q}^c_j}}{m^2_{\tilde{q}_i}-m^2_{\tilde{q}^c_j}}
\ln \frac{m^2_{\tilde{q}_i}}{m^2_{\tilde{q}^c_j}})\tilde{m}_j^qv_q~,
\end{equation}
where $\tilde{q}$ stands for the up squarks to Eq. (4) and the down squarks
to Eq. (5).\par
\vspace{1.0cm}
   To require that there is a further mass hierarchy between the second
and the first family, we assume that the right-handed squarks are degenerate.
In general, by including the $\delta M$ (6), the quark mass matrix would
become rank-three, the two light up quarks or the two light down quarks
would have  masses with the same order of magnitude.  By the assumption of
the degeneracy of the right-handed squarks, that is
$m^2_{\tilde{q}^c_j}=m^2_{\tilde{q}^c}$ (for $j$=1, 2, 3), the $\delta M$
matrix (6) can be written as
\begin{equation}
(\delta M)_{ij}=f_i\tilde{m}_j^q~.
\end{equation}
$f_i$ is a function of $m_{\tilde{g}}$, $m^2_{\tilde{q}^c}$ and
$m^2_{\tilde{q}_i}$.  It is for the left-handed squark masses we assume the
${\rm Z_{3Q}}$ symmetry is broken, that is
$m^2_{\tilde{q}_i}\neq m^2_{\tilde{q}_j}$ for $i\neq j$, otherwise the
light quarks will be still massless.  Therefore $f_i$ depends on the flavor
of the left-handed quarks.  With the above "factorization" (7), the form
of the mass matrix for both the up quarks and the down quarks is
\begin{equation}
M+\delta M=\left( \begin{array}{ccc}
a+f_1\tilde{m}_1&b+f_1\tilde{m}_2&c+f_1\tilde{m}_3\\
a+f_2\tilde{m}_1&b+f_2\tilde{m}_2&c+f_2\tilde{m}_3\\
a+f_3\tilde{m}_1&b+f_3\tilde{m}_2&c+f_3\tilde{m}_3\\ \end{array} \right)~,
\end{equation}
where $a$, $b$, $c$ denote the tree level masses in Eqs. (4) and (5).  It
is easy to show that the rank of the mass matrix (8) is two.  Thus at this
stage, only the second family acquires masses.  The first family remains
massless.  A hierarchy between the second family and the first family
emerges.\par
\vspace{1.0cm}
   The order of magnitude of the masses of the second family can be
understood naturally.  From Eq. (6), we see that the charm quark to the
strange quark mass ratio $m_c/m_s$ is mainly determined by the ratio
$(\tilde{m}^uv_u)/(\tilde{m}^dv_d)$ if there is no significant difference
between the masses of the squarks with same chirality.  The ratio
$(\tilde{m}^uv_u)/(\tilde{m}^dv_d)$ can be $m_t/m_b\sim O(10)$.  Therefore
the large ratio of $m_c/m_s$ can be considered as a result of $m_t/m_b$.
It should also be noted that in Eq. (6), we have neglected the other
neutral gauginos, photino and Zino, because their effects are rather small
compared with that of gluinos for the following two reasons.  One is that
$\alpha_s$ is large, $\alpha_s/\alpha\sim O(10)$; another is that the
number of gluinos is 8 which is also large.  Hence the contribution of
gluinos is nearly two orders of magnitude larger than that of photino or
Zino.  The radiative mass generation picture of quarks discussed above
can be consistent with that of leptons of Ref. [3] where it is
the electron mass that is generated at the one-loop level by exchanging
photino and Zino.  The fact that the strange quark is two orders of
magnitude heavier than the electron is thus explainable.  A numerical
illustration will be given in Sec. 5.\par
\vspace{1.0cm}
   From the mass matrices with the form given by Eq. (8), the quark mixing
matrix can be obtained straightforwardly.  The mass eigenvalues of the
three families are written as follows,
\begin{equation}
\begin{array}{rcl}
m_3 &\simeq& \sqrt{3} (|a|^2+|b|^2+|c|^2)^{1/2}~,\\
m_2 &\simeq& \frac{1}{\sqrt{3}}(|\alpha\tilde{m}_2-\beta\tilde{m}_1|^2
+|\beta\tilde{m}_3-\gamma\tilde{m}_2|^2
+|\gamma\tilde{m}_1-\alpha\tilde{m}_3|^2)^{1/2}\\
&&\cdot(|f_1-f_2|^2+|f_2-f_3|^2+|f_3-f_1|^2)^{1/2}~,\\
m_1 &   =  & 0~,
\end{array}
\end{equation}
with $\alpha=\frac{a}{\sqrt{|a|^2+|b|^2+|c|^2}}$,
$\beta=\frac{b}{\sqrt{|a|^2+|b|^2+|c|^2}}$,
$\gamma=\frac{c}{\sqrt{|a|^2+|b|^2+|c|^2}}$.
Without loss of generality, we simply take $a=b=c$, $\tilde{m}_i=f_im_0$,
and neglect $m_c/m_t$ in the calculation.  In this case, the
mixing matrix can be easily obtained,
\begin{equation}
V_{\rm mixing}\simeq\left( \begin{array}{ccc}
1&0&0\\
\displaystyle 0&1&-\frac{f\sqrt{m_sm_0}}{\sqrt{3}m_b}\\
\displaystyle 0&\frac{f\sqrt{m_sm_0}}{\sqrt{3}m_b}&1\\ \end{array}
\right)~,\\[3mm]
\end{equation}
with $f=\sum_{i}^{}f_i$.  We note that if $f_1=f_2=0$, the mass matrices
further reduce
to the form assumed in Ref. [8], in which the quark mixing matrix element
$V_{cb}$ can be consistent with the experimental value after including the
$m_c/m_t$ correction.  As for the Cabbibo angle, it is determined by the
mass ratio of the first to second family, which is nonzero only
when the nonvanishing quark masses of the first family are introduced, and
will not be discussed in this paper.\par
\vspace{1.0cm}
   It should be noted that all the discussions about the quark mass matrix
as well as the quark mixings in this section are actually irrelevant to the
R-parity violating interactions, because the ${\rm Z_{3Q}}$ symmetry has
made these interactions have no effects.  The R-parity violation is
introduced for the consistency in the framework of the quark sector and the
lepton sector in Ref. [3].  It will be important in the discussion of CP
violation in the next section.\par
\vspace{1.5cm}
{\large \bf 4. CP violation}\par
\vspace{1.0cm}
   In general, there are several possible origins of CP violation within
the framework of supersymmetry.  The first one is the complex Yukawa
interactions, as in the Standard Model.  In this model, the complex Yukawa
couplings cannot be the source of CP violation because of the discrete
family symmetry introduced, the phases of the Yukawa couplings can all be
absorbed by the redefinition of the quark fields.  The second possible
origin is the phases of the soft breaking terms.  Such an origin is a
specific feature of supersymmetry theory [9].  These phases would in turn
enter the quark mixing matrix through the radiative mass generation
mechanism (6).  However, the experimental data of the neutron electric
dipole moment constrains most of these phases to be very small.  Although
there is still no complete analysis about the phases of the R-parity
violating soft terms, we expect that they are equally small.  Therefore,
the soft breaking terms are also not the source of the observed CP
violation.  The third origin of CP violation lies in the complex VEVs of
the sneutrino fields.  This is a special feature of this  model.  As we
have explained in Sec. 2, the VEVs of the Higgs fields and the sneutrino
fields are nonvanishing.  At this point, this model is similar to the
multi-Higgs doublets models [10, 11] in which CP can break spontaneously,
the VEVs are complex generally [12].\par
\vspace{1.0cm}
   The CP violation occurs through the sneutrino exchange in our model
due to the complex VEVs of the sneutrino fields.  The magnitude of these
VEVs is very small compared to the Higgs VEVs.  From the superpotential
(3), it can be seen that, in general, the couplings of the R-parity
violating terms are not diagonal in the basis in which the Yukawa
couplings are diagonal.  They result in FCNC at tree level for
the down quarks.  To avoid too large FCNC, we assume that these couplings
are very small compared with the Yukawa couplings.  Therefore our model
can be viewed as a kind of minimal supersymmetric standard model but with
some small deviations.  This case is also very similar to that studied in
Ref. [13] in which the flavor changing couplings in the two Higgs doublets
model are assumed to be small so as to be treated perturbatively.  Some
conclusions there can be applied here.  Now let us consider the
${\rm K^0-\bar{K}^0}$ system which gives one of the most severe restrictions
on the FCNC.  By taking the FCNC couplings as expansion parameters, then
the lowest order $\Delta S =2$ effective Lagrangian induced by the tree
level sneutrino exchange is obtained approximately as
\begin{equation}
L_{\Delta S=2}^{\not R}\simeq\frac{\lambda'^2}{m_{\tilde{\nu}}^2}
(\bar{d}\gamma_5s)^2+{\rm h.c.}~,
\end{equation}
where $m_{\tilde{\nu}}$ is the typical sneutrino mass and $\lambda'$
stands for the FCNC couplings in Eq. (3).  To meet the experimental results,
the contribution of Eq. (11) to the ${\rm K_L-K_S}$ mass splitting
$\Delta m_{\rm K}$ should be smaller than that of the box diagram through the
$W$ exchange.  At this stage, CP is still conserving if
$\lambda'$ is real.  Because of the complex sneutrino VEVs which are also
small, CP violation is introduced at the next order of $\lambda'$ [13].
Compared to the Lagrangian (11), the CP violation is suppressed by a
factor of $\lambda'\sum_{i}^{}v_i/m_s$,
\begin{equation}
L_{\Delta S=2}^{\not{CP}}\simeq\frac{\lambda'^2}{m_{\tilde{\nu}}^2}
\frac{\lambda'\sum_{i}^{}v_i}{m_s}(\bar{d}\gamma_5s)^2+{\rm h.c.}~.
\end{equation}
The effective Lagrangian above explains the CP violation in the
${\rm K_L-K_S}$ mixing.  The suppression guarantees that the R-parity
violating couplings are not extremely small or the sneutrinos not
extremely heavy.\par
\vspace{1.0cm}
   Numerically, we take $\lambda'\sim 10^{-4}-10^{-5}$, $v_i\sim 10$ GeV
and $m_{\tilde{\nu}}\sim (300-1000)$ GeV.  This choice of the parameters
is consistent with that in the lepton case [3] except that the R-parity
violating couplings in the quark sector are two orders of magnitude lower
than that in the lepton sector.  FCNC is thus small enough to be
consistent with experiments.  Approximately, Eq. (11) contributes $1/10-1$
of the ${\rm K_L-K_S}$ mass difference.  The above choice also justifies
the suppression of CP violation,
$\lambda'\sum_{i}^{}v_i/m_s\sim 10^{-2}-10^{-3}$.  Hence the CP violation
parameter $\epsilon$ in the ${\rm K_L-K_S}$ mixing can be in agreement with
the observation, $\epsilon\sim 10^{-3}$.\par
\vspace{1.0cm}
   This mechanism for CP violation is superweak.  The other CP violation
parameter $\epsilon'/\epsilon$ is too small to be observable.  Because
there is no tree level FCNC for the up quarks, ${\rm D^0-\bar{D}^0}$
system has no CP violation in our case.  Its implications on the neutron
electric dipole moment and the CP violating effects in the
${\rm B^0-\bar{B}^0}$ system need to be studied further.  It should be
also remarked here that if there is no CP violation in the soft breaking
sector, the mechanism proposed here is a spontaneous one for CP
violation.\par
\vspace{1.5cm}
{\large \bf 5. Supersymmetric FCNC}\par
\vspace{1.0cm}
   Last section has considered the tree level FCNC of the R-parity
violating interactions.  In this section, we briefly discuss the FCNC
induced by the superpartners of the ordinary particles in loops. The
experimental data on FCNC put severe constraints on the extensions of
the Standard Model.  It is well-known that the superpartners would
result in unacceptable large FCNC through loop graphs [14] unless some
degeneracy conditions are imposed on the squarks.  Usually these conditions
are\par
\vspace{0.5cm}
   (i) The squark mass-squared matrix corresponding to the left-handed
quarks is proportional to unit matrix.  That is
$m^2_{\tilde{q}_1}=m^2_{\tilde{q}_2}=m^2_{\tilde{q}_3}$.\par
\vspace{0.5cm}
   (ii) The squark mass-squared matrix corresponding to the right-handed
quarks is proportional to unit matrix.  That is
$m^2_{\tilde{q}_1^c}=m^2_{\tilde{q}_2^c}=m^2_{\tilde{q}_3^c}$.\par
\vspace{0.5cm}
   (iii) The squark mass-squared matrix corresponding to the mixing
associated with different chiralities is proportional to the relevant
quark mass matrix.  That is $\tilde{m}^qv_q=\tilde{m}^0M^q$, where
$\tilde{m}^0$ is a common mass parameter.\par
\vspace{0.5cm}
   In our case, it is easy to require that the conditions (ii) and (iii)
are satisfied.  Actually, they are what have been assumed in Sec. 3 in
obtaining the hierarchy between the second and the first family.  While
this is a good news for us, however, the condition (i) cannot be satisfied,
otherwise the second family will be massless.\par
\vspace{1.0cm}
   The radiative generation of the quark masses for the second family
requires some violation to the condition (i).  Of course, this violation
should be small enough to be consistent with experiments.  In Sec. 3, the
generation mechanism of the masses for the second family has implies that
the left-handed squark masses are not universal.  Their mass can be
expressed as
\begin{equation}
m^2_{\tilde{q}_i}=m^2_{\tilde{q}}+\delta m^2_{\tilde{q}_i}~,
\end{equation}
where $\delta m^2_{\tilde{q}_i}$ ($i$=1, 2, 3) denotes the small deviation
to the universal mass limit $m^2_{\tilde{q}}$.  By choosing that the
gluino mass $m_{\tilde{g}}$, the right-handed squark mass $m_{\tilde{q}^c}$
and the universal left-handed squark mass $m_{\tilde{q}}$ are equal, Eq.
(6) is expressed simply as follows,
\begin{equation}
(\delta M)_{ij}\simeq\frac{\alpha_s}{\pi}\frac{\delta m^2_{\tilde{q}_i}}
{m^2_{\tilde{q}}}\frac{\tilde{m}^q_jv_q}{m_{\tilde{q}}}~.
\end{equation}
Experimentally, with the requirement that the supersymmetric contributions
to the $\Delta m_{\rm K}$ and $\Delta m_{\rm B}$ are smaller than the
measured values,
the analysis of Ref. [14] gives the following upper limit for
$\delta m^2_{\tilde{q}_i}/m^2_{\tilde{q}}$,
\begin{equation}
\frac{\delta m^2_{\tilde{q}_i}}{m^2_{\tilde{q}}}<0.1
\frac{m_{\tilde{q}}}{1{\rm TeV}}~.
\end{equation}
Without being contrary to the experiments on FCNC, we take
$m_{\tilde{q}}=m_{\tilde{q}^c}=m_{\tilde{g}}=300$ GeV,
$\delta m^2_{\tilde{q}_i}/m^2_{\tilde{q}}\simeq 0.01$ as a numerical
illustration which is also consistent with the analysis of
$b\rightarrow s\gamma$ [14].  In this case, $\tan\beta\simeq 2.2$,
$\tilde{m}^u\simeq 5$ TeV and $\tilde{m}^d\simeq 1$ TeV can result in the
realistic values for the masses of the charm quark and the strange
quark.\par
\vspace{1.0cm}
   Some remarks should be made in the following.\par
\vspace{0.5cm}
   (a) Because the CP violation mechanism in our framework, which has been
described in Sec. 4, is different from the one of Ref. [14], the constraints
given in Ref. [14] from the analysis of CP violation are not valid here.\par
\vspace{0.5cm}
   (b) It is possible to imagine that there is also a small violation to
the condition (ii).  In this model, such violation is severely constrained
by the masses of the first family of quarks.  By assuming some deviations
$\delta m^2_{\tilde{q}_j^c}$ to the condition (ii), it can be seen from
Eq. (6) that the quark mass matrix $M+\delta M$ will then become rank-three.
Hence the masses of the first family are produced.  By requiring that the
produced masses of the first family are smaller than the measured values,
it is straightforward to obtain the constraint
$\delta m^2_{\tilde{q}_j^c}/m^2_{\tilde{q}}<10^{-4}$ which is about
two orders of magnitude lower than that given by FCNC [14].\par
\vspace{0.5cm}
   (c) Although in this framework the parameters of the soft breaking terms
are arbitrary and just fitted to experiments, some of them are as large as
several TeV.  This should be explained in some underlying theory which
describes the supersymmetry breaking.\par
\vspace{1.5cm}
{\large \bf 6. Summary and discussion}\par
\vspace{1.0cm}
   In this paper, we have described a pattern of quark mass hierarchy and
CP violation within the framework of low energy supersymmetry.  By assuming
some discrete symmetry among the three families, the quarks of the third
family obtain masses at the tree level.  The second family obtains
masses radiatively at the one-loop level due to the
soft breaking of the family symmetry.  At this level, the first family
remains massless by some degeneracy conditions of the squarks.  As a
result of R-parity violation, the sneutrino VEVs
are nonvanishing.  CP violation occurs through the superweak sneutrino
exchange.  The above picture is consistent with the experiments on
FCNC.\par
\vspace{1.0cm}
   It can be seen that the understanding of both the fermion masses and CP
violation in this work depends essentially on the supersymmetry.  Usually
the researches on R-parity violation include two aspects, one is the
trilinear interactions; the other is the nonvanishing sneutrino VEVs.
These two aspects were discussed separately before [1].  We have combined
them so as to discuss the fermion mass and CP violation problems.  Its
implications on astrophysics should be studied further [15].  The
violation to the discrete symmetry has been introduced in the
supersymmetric soft breaking sector explicitly.  While this needs further
explanation, it may avoid the cosmologically domain wall problem.\par
\vspace{1.0cm}
   Although logically the discussions on the quark mass hierarchy, the CP
violation mechanism and the lepton mass hierarchy in Ref. [3] could be
separate stories, we would like to unite them together.  This will give
some more definite results.  In addition to the prediction of the
Marjorana neutrino masses being at the 1 eV range in Ref. [3], the CP
violation parameter $\epsilon'/\epsilon$ is predicted to be too small to
be observable.  More predictions need further analysis which should
include introducing the masses of the first family of quarks and
calculating the CP violation in detail.\par
\vspace{1.0cm}
   There is a hierarchy between the Higgs VEVs and the sneutrino VEVs,
despite that this hierarchy is not large numerically.  This in turn
would give some
information about the scalar potential, in which the parameters may also
have some hierarchy.  In particular, the couplings of the R-parity
violating interactions are much more smaller than the Yukawa couplings.
All of these need natural explanations from some underlying theory which
is under our consideration.\par
\vspace{1.0cm}
   Finally we would like to point out a possible scenario of the mass
generation of the first family.  In the discussion about the quark sector
in Sec. 2, two cyclic discrete symmetry groups
${\rm Z_{3L}}\times{\rm Z_{3Q}}$ have been assumed.  While they
are introduced to avoid some confusion, we can imagine that they are
replaced by the diagonal subgroup of them from the begining, which is the
cyclic ${\rm Z_3}$ symmetry among the left-handed SU(2) doublets
of the three families including both the leptons and the quarks.
Such a case is practically the same as what we have discussed in this
paper, because the quantity $\lambda'v_i$ appeared in Eq. (5) has been
taken to be not only much smaller than the masses of the third family, but
also smaller than the masses of the second family.  However, this quantity
will produce a mass to the down quark of the first family in this case,
and its value can be several MeV numerically.  This scenario also justifies
the CP violation mechanism discussed in this paper.  In addition, the mass
of the up quark of the first family cannot be produced in this way.  This
gives us an explanation of the fact $m_d>m_u$, and even may bring us a
solution to the strong CP problem.\par
\vspace{2.0cm}
\begin{center}
{\bf Acknowlegement}
\end{center}
\vspace{1.0cm}
   We would like to thank D.S. Du for his encouragements and discussions
and Z.Z. Xing for helpful discussions.

\newpage

{\large \bf Figure caption}\\

Fig. 1.  Supersymmetric generation of the light quark masses, where
$\tilde{g}$ and $\tilde{q}$ denote the gluino and the squark, respectively.

\end{document}